# Nanoscale molecular electrochemical supercapacitors

Ritu Gupta, Ankur Malik, Vincent Vivier* and Prakash Chandra Mondal*

Department of Chemistry, Indian Institute of Technology Kanpur, Uttar Pradesh-208016, India

Sorbonne Université, CNRS, Laboratoire de Réactivité de Surface, 4 place Jussieu, Paris 75005 Cedex 05, France

E-mail: vincent.vivier@sorbonne-universite.fr (VV); pcmondal@iitk.ac.in (PCM)

**Abstract**

Due to the shorter channel length allowing faster ion/charge movement, nanoscale molecular thin films can be attractive electronic components for next-generation high-performing energy storage devices. However, controlling chemical functionalization and achieving stable electrode-molecule interfaces at the nanoscale via covalent functionalization for low-voltage operational, ultrafast charging/discharging remains a challenge. Herein, we present a simple, controllable, scalable, low-cost, and versatile electrochemical grafting approach to modulate chemical and electronic properties of graphite rods (GRs) that are extracted from low-cost EVEREADY cells (1.5 US $ for 10 cells of 1.5 V). On the ANT-modified GR (ANT/GR), the total capacitance unveils 350-fold enhancement as compared to an unmodified GR tested with 0.1 M $H_2SO_4$ electrolyte ensured by both potentiostatic and galvanostatic measurements. Such enhancement in capacitance is attributed to the contribution from the electrical double layer and Faradaic charge transfer. Due to higher conductivity, anthracene molecular layers possess more azo groups (-N=N-) over pyrene, and naphthalene molecular films during the electrochemical grafting, which is key to capacitance improvements. The ultra-low-loading nanofilms expose high surface area leading to extremely high energy density. The nanoscale molecular films (~ 23 nm thickness) show exceptional galvanostatic charge-discharge cycling stability (10,000) that operates at low potential. Electrochemical impedance spectroscopy was performed along with the DC measurements to unravel in-depth charge storage performances. Electrochemically grafted molecular films on GR show excellent balance in capacitance and electrical conductivity, high diffusion coefficient toward ferrocene, and can easily be synthesized in good yield on rigid to flexible electrodes. The polyaromatic-based molecular nano-films serve as a remarkable benchmark that holds the potential to replace the traditional inorganic metal oxide-based electrodes and build up next-generation molecular power banks.

# 1. Introduction

The increasing global population and depletion of fossil fuels have raised significant demands for technological gear, especially in portable, and lightweight energy storage systems.[1,2] To meet ever-expanding energy storage requirements, organic supercapacitors devices have been expanding but it is vital to elevate the energy density and power density. Excellent ability to synthesize, light-weight, low-production cost, tunable electrical conductivity, porosity, and sustainable organic electrodes easy fabrication under ambient conditions make organic supercapacitors for advanced technological solutions for storing electrical energies in different forms.[3–6] Thanks to the heteroatoms (N, P, O, S) due to their relatively larger electronegativity than carbon, thus increase surface wettability, alter the Fermi energy ($E_F$), and enhance capacitive performance upon functionalization on carbon electrodes.[7,8] Generally, the energy density of OSC can be improved either by broadening the operational voltage window (which also depends on the nature of the electrolyte) or increasing the specific capacitance of electrode materials by combining electrical double-layer capacitance (EDLC) and pseudocapacitance.[9] Over the last decades, either organic or inorganic "nanomaterials" have played significant roles in electrical energy storage applications.[10,11] In addition to the advantages of high-surface-area, the critical size of energy storing materials shrinks to nanoscale dimension (~1-100 nm length) thus diffusion length gets reduced compared to the bulk, which fastens the charging-discharging, increases the cycle life, and lowers the operational bias, thus enhancing overall supercapacitor performances.[12] All these advantages clearly illustrate that the nanoscale structural design and chemistry of modified organic electrodes may provide a new generation of devices that can address the theoretical limit for electrochemical storage and can emancipate electrical energy efficiently and rapidly.[13] Among the promising candidates, graphite rod (GR) has been extensively exploited in dry cells and electrochemical supercapacitors due to its high double-layer capacitance and lower cost; however, it suffers from very low energy density and lower cyclic stability.[14–16]

In conjunction with the emergence of these nanostructured carbon materials, there has been a surge of interest in controlling surface physical and chemical properties by covalent surface modifications.[17–19] Most covalent surface modifications do not deliver spatial control in the nanoscale dimension, which is highly desirable for developing an efficient charge storage system. Such covalent surface functionalization is challenging and sometimes rely on high-energy reagents and harsh reaction conditions that making uncontrollable nanostructures.[4,20–22] To address this issue, electrochemical reduction of aryl diazonium is the most efficient approach as it can create strong, stable, robust covalent bonding between the molecule and the substrates, especially with carbon for multi-purpose usages.[23–26] Though surface modification of carbon materials by spontaneous reduction of diazonium has already been studied in energy storage.[27,28] However, the technique does not provide a controlled, homogeneous nanoscale multilayer film and requires longer time. For that reason, we have opted a simple, less time-consuming bottom-up approach to modify graphite rod (GR) with high surface area, and nitrogen-rich graphene structures formed via electrochemical reduction of diazonium salts of three different compounds, 1-amino anthracene (ANT), naphthalene (NAPH) and pyrene (PYR) and electrochemical capacitance performances compared with each molecular layer. So far, very few reports are made with E-Chem modification in energy storage. The ANT-modified GR shows not only high charge storage density but also long cycle life of up to $10^4$. In addition, we have reused the used graphite rods from commercially available EVEREADY low-cost cells for electrochemical supercapacitor studies upon growing nanoscale molecular films, which also reveals excellent charge storage compared to the unmodified one. The present study comprises low-cost materials that can be easily scaled up to fabricate high-capacity energy systems for practical applications in molecular energy storage at the nanoscale and the concept of 'waste to energy' can be viable.

## 2. Result and discussion

The synthesis and characterization of respective diazonium salts are discussed in supporting information (**Fig. S1-S4, Table S1**) which are used for the growth of molecular layers on graphite rod (GR) by electrochemical reduction method. Being highly reactive, generated at the electrode surface, aryl radicals covalently bonded graphite rod used as working electrodes and extended with a repeated number of cyclic voltammograms (CVs) scans, keeping potential window unchanged. Continuous binding of (for instances, ANT) radicals maintains the growth of anthracene molecular layers along with azo groups (−N=N−) as bridges. However, though previous studies on E-Chem grafted molecular layers utilized for hybrid supercapacitors studies, but none of the work highlights such unexpected azo formation during E-Chem grafting. Non-radical formation of azo bridges between layers is one of the main assumptions to explain the layer growth other than radical polymerization (**Fig 1a,** inset shows graphitic rod). **Fig 1b** shows the potentiostatic cyclic voltammograms (CVs) scans (1-20 cycles) of electrochemical modification of GR substrate with 5 mM of ANT-D in acetonitrile recorded at 100 mV/s scan rate using 0.1 M *n*-tetrabutyl ammonium perchlorate (TBAP) as a supporting electrolyte. CV shows the two reduction peaks: a broad irreversible reduction peak at −0.25 V vs. Ag/AgNO$_3$ was observed during the first scan indicating the reduction of anthracene diazonium salts (ANT-D), and a higher cathodic peak indicating formation of ANT radical to ANT anion. To determine thickness of ANT layers, ANT-D was E-Chem grafted on indium tin oxide (ITO) substrate with the same electrochemical parameters used for GR modification. AFM image of ANT film and thickness of the film was evaluated by scratching the ANT films with a sharp needle and 23 ± 0.8 nm thickness was obtained from depth analysis (**Fig S5-S6**). Amorphous nature of the bare graphite rod and ANT-modified GR was analyzed by X-ray diffraction (XRD) studies (**Fig 1c, S7**). The diffraction peaks at 2θ values of 26.5, 42.4, 44.5, 54.5, 77.5, and 83.6° corresponding to the (002), (100), (101), (004), (103), and (110) planes of hexagonal graphite with P63/mmc space group (JCPDS 00-008-0415), were observed for both. Field-emission scanning electron microscopy (FE-SEM) image of bare GR shows micrometer-sized flake-like structures (**Fig S8**), while ANT/GR shows increased roughness, indicating growth of molecular films (**Fig 1d**). The Raman spectra of bare GR and ANT/GR show a strong D-band at 1351 cm$^{-1}$, a G-band at around 1583 cm$^{-1}$, and a strong 2D-band around 2705 cm$^{-1}$ (**Fig 2a, S9**). The D band is observed due to the breathing mode of *k*-point phonons of $A_{1g}$ symmetry of the porous carbon, which also reveals the presence of defect sites, impurities, and disorder in the structure.[29,30] The $I_D/I_G$ ratio was 0.12 and 0.36 for bare GR and ANT/GR, respectively, indicating increased disorders in ANT/GR. The G band represents the $E_{2g}$ phonon of $sp^2$ carbon atoms and corresponds to the ordered $sp^2$ hybridized graphene network.[31] The 2D peak was observed at about 2700 cm$^{-1}$, caused by the second-order vibration of two phonons with opposite wavevectors.[31] The expended and deconvoluted x-ray photoelectron spectra of the C1s peak in bare GR were deconvoluted into two peaks, one at 284.6 eV for C═C and the other at 285.6 eV, corresponding to C─C (**Fig 2b**).

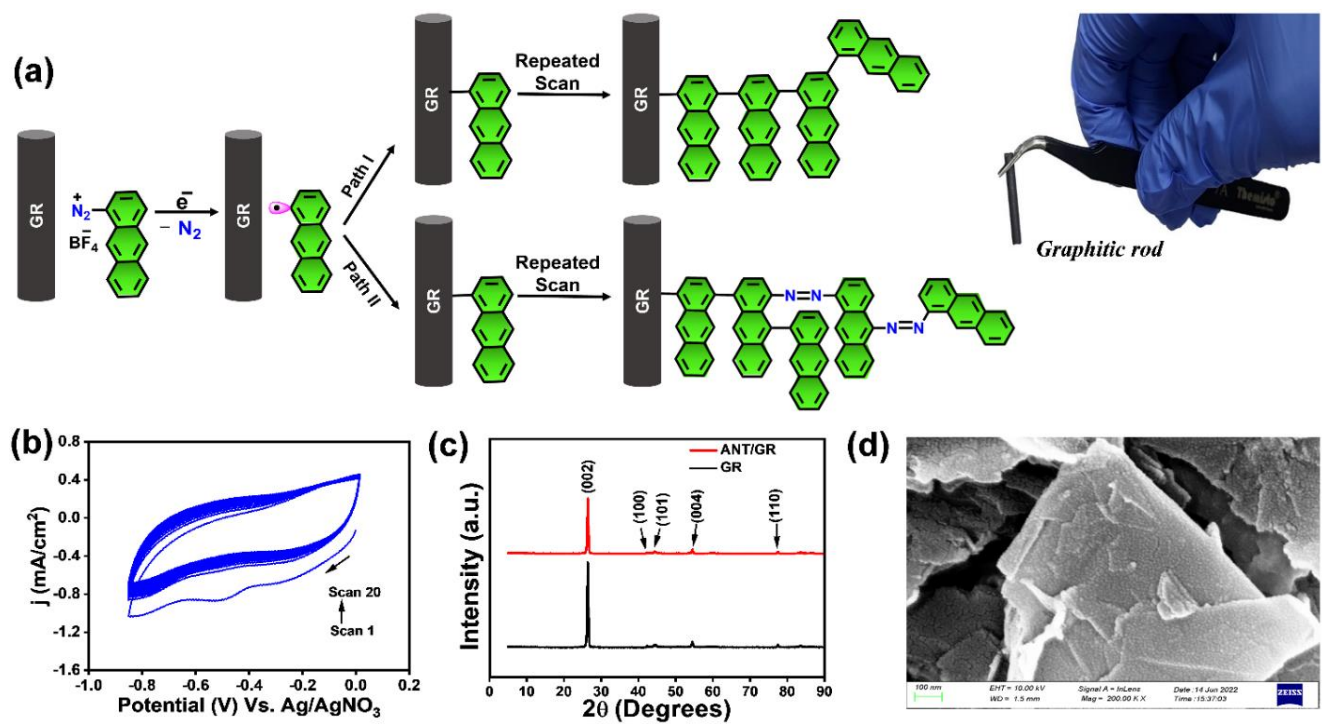

**Fig 1.** (a) Schematic description of proposed pathways of E-chem grafting of ANT-D on graphite rod (GR) leading to multilayer formations (digital image of GR is shown in right corner). (b) Cyclic voltammograms of E-chem grafting of ANT-D of 5 mM in ACN with 0.1 M TBAP recorded at 100 mV/s during 20 scans. (c) X-ray diffraction pattern of bare GR (black) and ANT-modified films on GR electrode (red). (d) FE-SEM image of ANT/GR electrode.

The expended and fitted XPS spectra of C1s of ANT/GR show three contributions at 284.5 eV, 284.9 eV, and 285.7 eV for C═C, C─C, and C─N, respectively (**Fig 2c, S10**) which are well-correlated with the previous report.[32] The presence of nitrogen was observed in ANT/GR; but absent in bare GR (**Fig S11**). In ANT/GR, two broad peaks in the range of 398 eV to 402 eV were obtained, which were further deconvoluted into five peaks at 398.6 eV, 399.5 eV, 400.4 eV, 402.2 eV, and 402.8 eV suggesting the presence of five types of Ns at pyridinic N, pyrolytic N, graphitic N, quaternary N, and N─N, respectively,[33] thus confirming the stacking of ANT rings with azo bridges and forming nitrogen-rich GR structures. The binding energy of elements in both bare GR and ANT/GR electrodes are given in **Table S2**. The BET surface area was evaluated by measuring the $N_2$ adsorption-desorption isotherm, and pore volume was calculated using the DFT simulation. The isotherms were of Type-III, indicating multilayer formation where the mesopores and macropores were not completely filled. The adsorption-desorption hysteresis was H3-type hysteresis suggesting a plate-like structure of bare GR and ANT/GR, which was also evident from FE-SEM and the existence of inkbottle and slit-shaped pores (**Fig S12**). The bare GR had a very low specific surface area of 2.05 m²/g and a pore volume of 0.004 cc/g, while it increases (2.41 m²/g) and pore volume (0.006 cc/g) with ANT/GR. A 20% increase in surface area and pore volume is potentially due to the growth of highly porous GR. **Fig S13** presents mesopores as most of the pore size distribution lies in the range of 10-50 nm with a slight contribution of macropores.

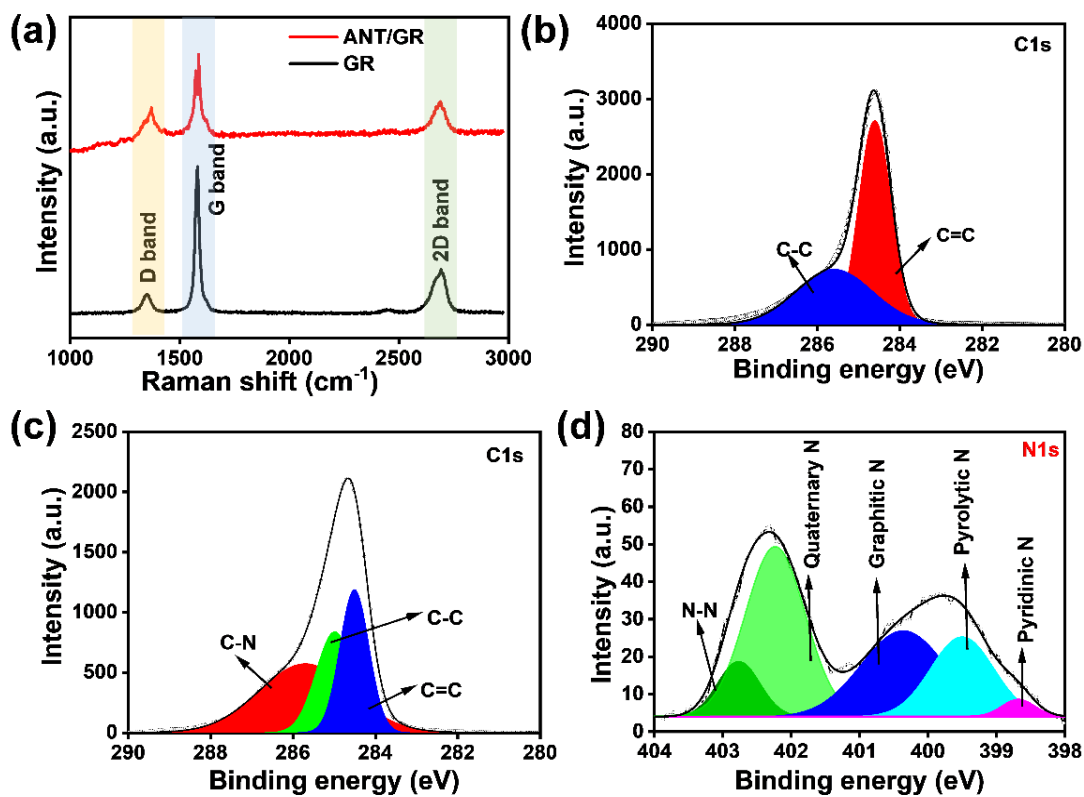

**Fig 2.** (a) Raman spectra of bare GR and ANT/GR. (b), (c) & (d) Deconvoluted and fitted XP spectra of C1s of bare GR, C1s of ANT/GR, and N1s of ANT/GR, respectively.

**Fig 3a** shows the cyclic voltammograms (CVs) of bare GR and ANT-modified GR recorded in 0.1 M H$_2$SO$_4$ (three electrodes set-up) at two different scan rates (CVs at other scan rates are shown in **Fig S14**). In contrast to bare GR, the ANT/GR electrodes exhibited a well-defined faradaic feature at about +0.31 V vs. Ag/AgCl, which can be ascribed to protonation/deprotonation of nitrogen present in the molecular films (similar to the redox phenomena), also capable of ultrafast charging/discharging. An increase in the peak-to-peak separation ($\Delta E_p$) was observed as a function of the scan rate, which is used to determine the apparent rate constant, $k_{app}$, using Laviron formalism.[34] At slower scan rates ($\leq$ 100 mV/s), $\Delta E_p$ remains constant in the 50-53 mV range but increases at higher scan rates. **Fig S15** shows the trumpet plot for ANT/GR in 0.1 M H$_2$SO$_4$ solution allowing the charge transfer coefficient, α, and 1−α calculated at 0.35 and 0.36, respectively. The $k_{app}$ of anodic reaction and cathodic reaction calculated at 14.9 ±0.8 and 14.0±0.6 s$^{-1}$, respectively, indicative of a fast-charge transfers process (see SI for detailed calculation) and nearly 10 times higher than the self-assembled organic redox active species.[35,36] Moreover, the reversible redox behavior of bare GR and ANT/GR electrodes was observed in presence of ferrocene, a well-known redox marker[37] (**Fig S16**). The diffusion coefficient, D related to the electrochemical nature of the charge transfer process, was deduced using the well-known Randles Sevcik equation (1).[38]

$$i_p = 2.695 \times 10^5 \, A \, D^{1/2} \, n^{3/2} \, v^{1/2} \, C \tag{1}$$

The anodic and cathodic currents were plotted as a function of the square root of the scan rate ($i_a$ vs. $v^{1/2}$) (**Fig S17**). From the slope of the plot, the diffusion coefficient (D) of ferrocene was calculated to be $4.2 \pm 0.18 \times 10^{-5}$ cm$^2$/s for bare GR and $5.6 \pm 0.44 \times 10^{-5}$ cm$^2$/s for ANT/GR electrodes, ensuring the facile formation of ferrocene-ferricinium species at the electrode surfaces.[37] The CVs of ANT-modified GR were recorded with varying electrolyte temperatures, and the activation energy ($E_a$) was calculated to be 10.5 ($\pm 0.4$) kJ/mol (**Fig S18**). The modified GR electrode show excellent thermal stability, heated at 75°C for 12 h (**Fig S19**). The total capacitance was further determined and compared for both systems by integrating the surface area obtained by the CVs in 0.1 M $H_2SO_4$ electrolyte and compared with naphthalene (NAPH) and pyrene (PYR) films (**Fig S20-21**). Bare GR shows electrochemical double layer capacitance (EDLC) of $24 \pm 1.36$ µF cm$^{-2}$ measured at a scan rate of 500 mV/s, which agrees well with typical carbon materials.[39] However, the ANT/GR electrode exhibits higher total areal capacitances with a well-defined faradaic feature at ~ 0.31 V vs. Ag/AgCl, thus illustrating the pseudo-capacitive behavior of the interface. As confirmed by XPS spectra, nitrogen of azo groups create several electroactive sites for pseudocapacitive reactions and protonation/deprotonation in response to electrochemical bias. A bar diagram is also shown in **Fig 3b** to compare the total areal capacitance of all modified active electrodes at a 500 mV/s scan rate showing the huge increase of the capacitance for the ANT/GR electrode. Generally, the total capacitance can be parted into two regimes, (i) EDLC and (ii) Faradaic reaction of active material. However, not all Faradaic reactions are pseudocapacitive in nature. Few are controlled by ion diffusion in the active electrode material including in the electrolytes.[40] Hence, it is necessary to evaluate the dominant Faradaic capacitive and non-capacitive contribution in order to impel an adequate explanation of the working performance of the active electrode. As shown in equation 2, the power law relationship can be used to distinguish the dominant charge storage process, where $i$ is current at a particular voltage, $v$ is scan rate, and a and b are constant.[41] The measured current fits linearly with the log ($i$) vs. log($v$) plot providing the b value from the slope. Values of b = 0.5 suggest, the direct relation of current with the square root of the scan rate and following the traditional diffusion-controlled approach. On the contrary, b = 1 implies the direct relation of current with the scan rate, which is the hallmark of a capacitor like charge storage behaviour.[41] CVs from low to high scan rates were used to calculate b values at the maxima anodic (**Fig 3c**) and cathodic regime (**Fig S24**). The linear fit in all three regions displayed a b value close to 1, indicating that the dominant charge transport mechanism in ANT/GR electrode is capacitor-like.

$$i(v) = a \, \vartheta^b \tag{2}$$

Further, capacitive contribution to the total current was evaluated by using the Dunn equation, shown below in equation 3, where $i$ is current at a fixed potential, $k_1 v$ and $k_2 v^{0.5}$ relates to capacitive and diffusion contributions to the total current.[41,42]

$$i\,(v) = k_1 v + k_2 v^{0.5} \qquad (3)$$

Thus, $k_1$ was calculated using the above equation at slow scan rates (5 mV/s to 50 mV/s), and 83% capacitive contribution was obtained at a 30 mV/s scan rate, as shown in **Fig 3d-e**. Further, the capacitive contribution was also calculated at different scan rates. With the increase in scan rates, the percentage of capacitive contribution also increases due to the slow diffusion of ions **(Fig 3f)**. The high capacitive contribution further claimed the fast redox reaction of ANT/GR electrode, which is highly desirable of pseudocapacitive materials.

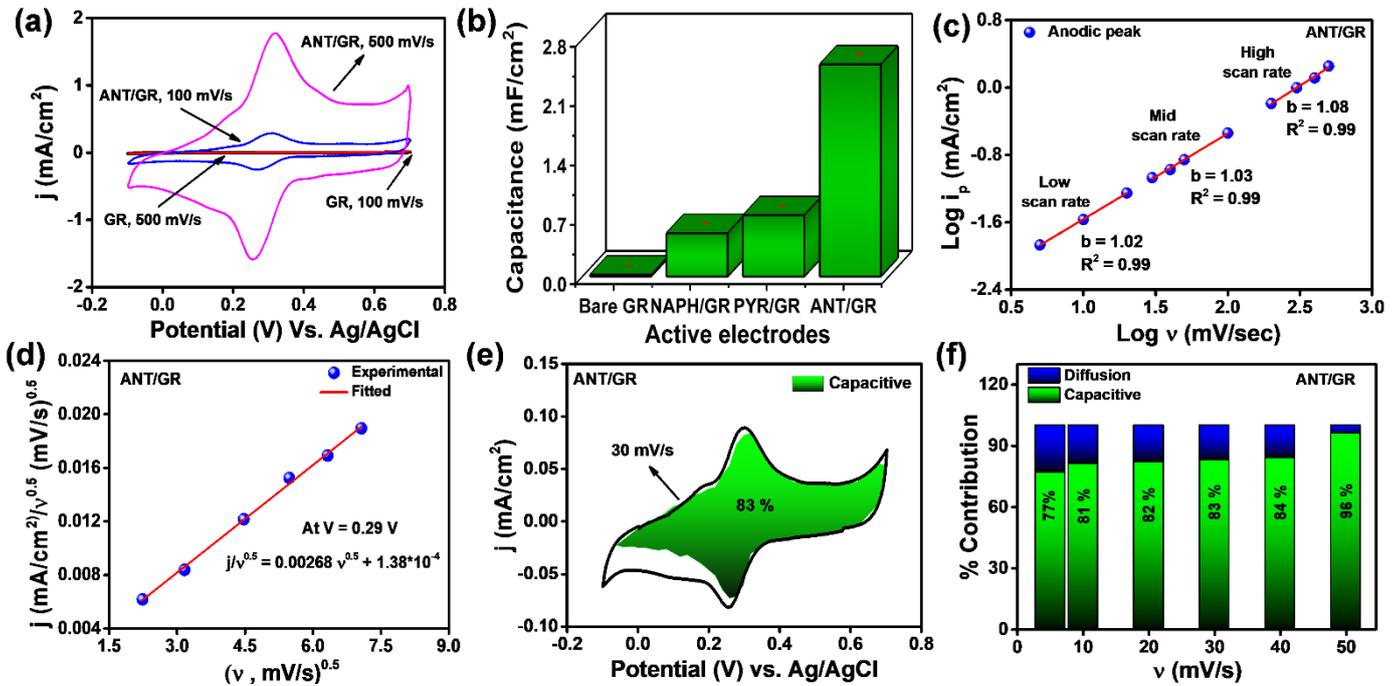

**Fig 3.** (a) Comparison of cyclic voltammograms of bare GR, ANT/GR in 0.1 M $H_2SO_4$ at 100 mV/s and 500 mV/s scan rates. (b) Bar diagram represnting total areal capacitance of bare and modified electrodes calculated from CV at 500 mV/s (with error bars in red color). (c) Plots of log ($i_p$) vs. log ($v$) from low to high scan rates for ANT/GR electrode in the anodic regime. (d) Plot of j ($mA/cm^2$)/$v^{0.5}$ $(mV/s)^{0.5}$ vs. ($v$, mV/s)$^{0.5}$ at 0.29 V. (e) CV at 30 mV/s showing capacitive (green) and diffusion-controlled (white) charge storage process. (f) Pseudocapacitive contributions of ANT/GR at different scan rates.

The capacitance performances were further evaluated from galvanostatic charge/discharge (GCD) cycling over potential ranges of −0.1 V vs. Ag/AgCl to + 0.7 V (vs. Ag/AgCl) at 30 µA cm$^{-2}$ for ten cycles (**Fig 4a-b**). Bare GR demonstrated a symmetrical and nearly triangular shape charge/discharge cycle, clearly indicative of EDLC feature. In contrast, a significant change in shape with a negligible potential drop, $V_{drop}$ (equivalent series resistance/ESR of 0.98 ± 0.02 mΩ.cm$^2$, **Fig. 4c**) was observed in ANT/GR which is the combined EDLC, and Faradaic contribution (protonation/deprotonation to nitrogen) in the capacitance, as depicted in **Fig 3b**. The duration of charging/discharging also increased after ANT modification, ascribed to the Faradaic process. A single charge/discharge cycle is shown in SI, illustrating different charge/discharge time scales before and after modifications (**Fig S25**). The ANT-GR electrode shows the highest capacitance of 11 mF cm$^{-2}$ at a lower current density of 10 µA cm$^{-2}$ and decreases with increasing applied current density (**Fig 4d-e**). However, this value differs from the CVs measurements, which is because the charging/discharging current in CVs was much higher than the GCD measurements. A single GCD cycle illustrates charging and discharging time for bare GR at 3 s and 4 s and ANT-modified GR at 52 s and 66 s (**Fig. S25**). The cycle life of ANT/GR was determined by performing $10^4$ GCD cycles at an applied current density of 1 mA cm$^{-2}$ (**Fig. S26**). It is noteworthy to note that ANT/GR exhibited excellent cycling performance up to $10^4$ GCD cycles and could hold larger charges for a longer time (**Fig 4f, S27**). Interestingly, a gradual increase in capacitance value with GCD cycling was observed, suggesting a greater number of active sites, confirmed by FE-SEM and BET analysis (**Fig. S28-30**). After $10^4$ GCD cycles, CVs were further performed, and there was a minor change in the shape of the CV; however, five times increment in capacitance value was noted (**Fig S31**).

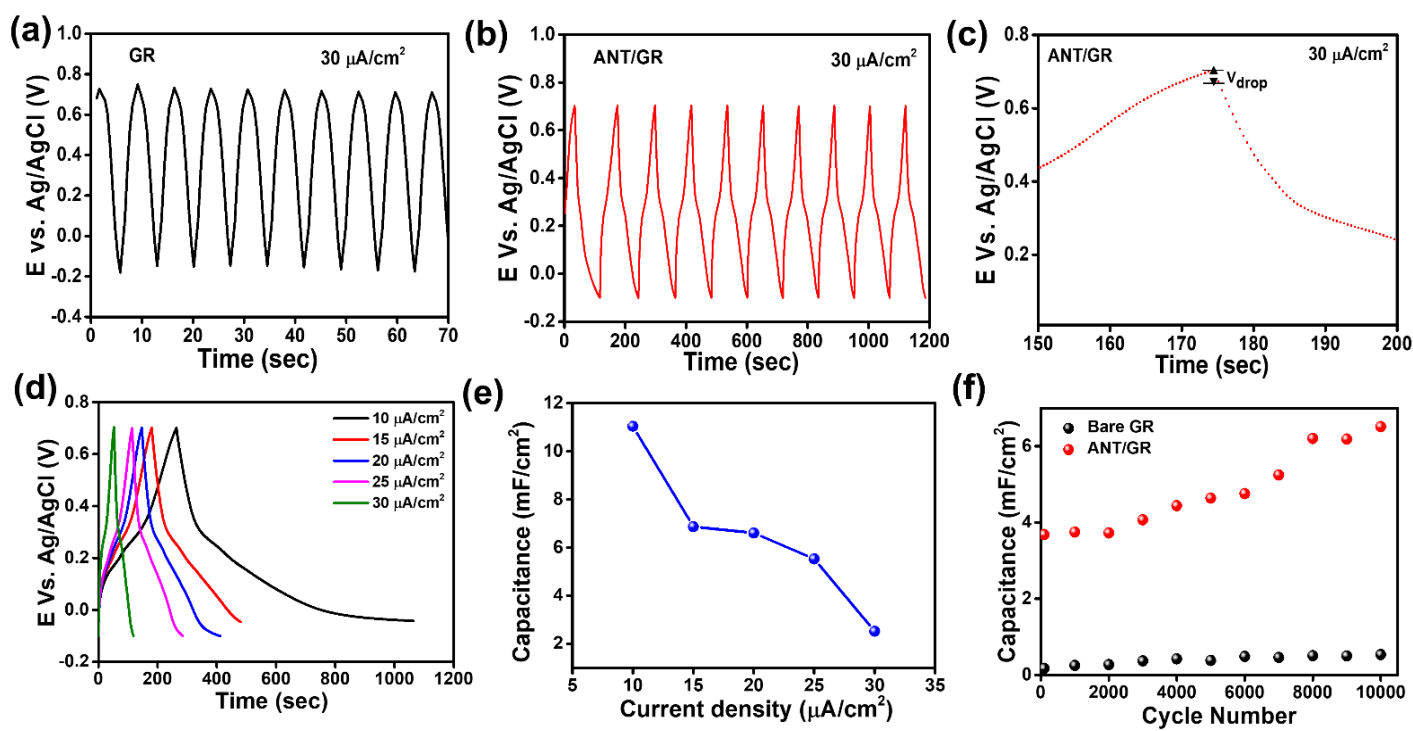

**Fig 4.** (a) Galvanostatic charging/discharging (GCD) curve of bare GR and (b) ANT/GR at 30 µA cm$^{-2}$. (c) A single GCD cycle of ANT/GR at 30 µA cm$^{-2}$ showing negligible potential drop, $V_{drop}$. (d) GCD plot of ANT/GR at different applied current densities. (e) corresponding capacitance values at different current densities. (f) Cycling performance of bare GR at the applied current density of 0.1 mA cm$^{-2}$ and ANT/GR at 1 mA cm$^{-2}$, to ensure the experiment time was the same for bare GR and ANT/GR, 0.1 mA cm$^{-2}$ current density was applied.

Areal energy density ($E_A$ in Wh cm$^{-2}$) and power density ($P_A$ in W cm$^{-2}$) are two crucial factors for electrochemical supercapacitors and are calculated from the experimental data at 30 µA cm$^{-2}$ (**section S16** in the SI). The bare GR exhibited an areal energy density of 0.01 µWh cm$^{-2}$ and a power density of 10.4 µW cm$^{-2}$. In contrast, the ANT-modified GR presented an areal energy density of 0.23 µWh cm$^{-2}$ and a power density of 12.21 µW cm$^{-2}$. Thus, compared to unmodified GR, ANT-modified GR showed 20 times increase in energy density; however, not much increase in power density was observed, demonstrating great potential for high energy density micro-supercapacitors.

Both the CV and GCD techniques are based on the DC-based measurements producing either total current or charges in the system but don't forecast frequency response which is crucial for understanding capacitance performance.[43] Electrochemical impedance diagrams obtained in frequency ranges of $10^{-1} - 10^4$ Hz for bare GR electrode and ANT-modified GR electrode before and after cycling are presented in **Fig. 5**. As expected, the bare electrode shows a constant phase element (CPE) (**Fig. 5a**) that can be better evidenced from the variation of the imaginary part of the impedance as a function of the frequency (**Fig. 5b**).[44,45] The slope of the curve is the opposite of the coefficient $\alpha$ of the CPE, whereas the determination of $Q$ is possible from the following relationship, $Q = \frac{-1}{Z_j \times (2\pi f)^\alpha} \times \sin\left(\frac{\alpha\pi}{2}\right)$

Then, using the Brug formula, [46,47] $C = Q^{1/\alpha} R_e^{(1-\alpha)/\alpha}$

in which $R_e$ is the electrolyte resistance, a capacitance of about 1.5 µF cm$^{-2}$ can be obtained. Such a value is consistent with a double layer formed at a graphite/electrolyte interface.[48] On the ANT-modified electrode (**Fig. 5c**), the shape of the impedance diagram remains unchanged, but the modulus is 10 times smaller, corresponding to an increase in the active surface area. The graphical analysis of this diagram is also consistent with capacitive behavior (**Fig. 5d**). After $10^4$ GCD cycles, the shape of the impedance diagram clearly shows two linear domains corresponding to two CPE behaviors, whereas the modulus is about 100 times smaller than what is measured on the as-prepared electrode (**Fig. 5e**). This can be explained by a rearrangement of the layer upon cycling and the formation of small pores in the thin film. This is confirmed by the graphical analysis presented in **Fig. 5f**, which distinctly exhibits two different linear domains for the determination of α, one in the high frequencies with a value close to 0.66 and a second one in the lowest frequency domain with a value close to 0.81. Such a value in high frequency is consistent with a porous system. However, it is not possible to go further in the determination of the size, depth, and diameter of the pores because all these parameters are linked.[49] Moreover, the analysis of the graphically determined $Q$-parameter value (**Fig. 5f**) shows that the capacitance value is in the range of a few mF cm$^{-2}$, in agreement with the CV, and GCD results, but the change of the active surface during cycling does not allow a better quantitative analysis. Finally, these diagrams were obtained at Faradaic process at +0.3 V vs. Ag/AgCl that is superimposed on the capacitive response. Extrapolation of the diagram presented in **Fig. 5e** allows us to determine a charge transfer resistance of about 100 Ω. This value can only be obtained at very low frequencies because of the parallel behavior of resistance with a double layer capacitance of electrode of a very high value. Furthermore, if we relate this value to a unit area, the analysis of impedance diagrams shows that there is an increase of a factor of 100 to 1000 between a bare electrode and an electrode after cycling. We deduce that the charge transfer resistance is small, which is characteristic of a Nerstian system, again in agreement with the cyclic voltammetry results.

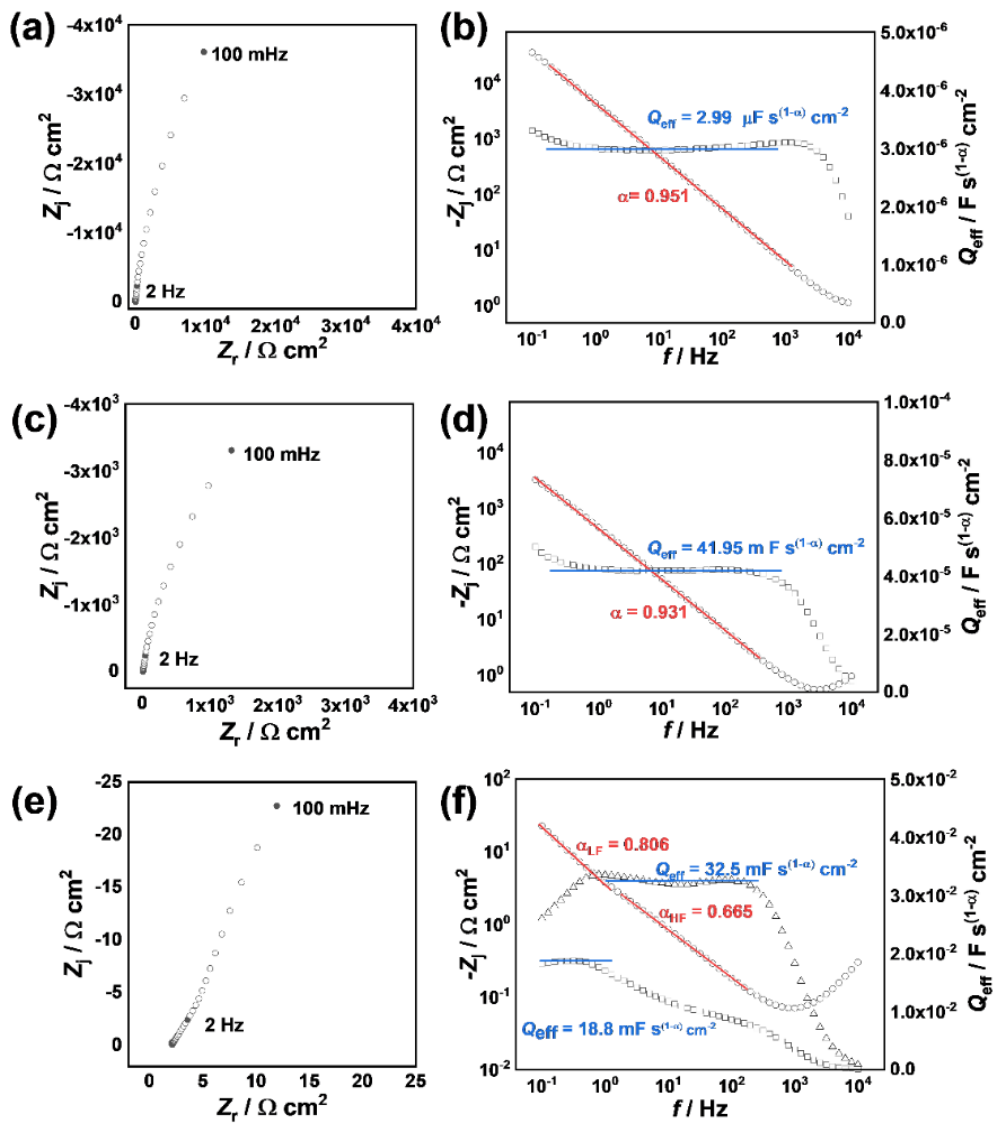

**Fig 5.** Electrochemical impedance responses and graphical analysis for the determination CPE parameters on (a-b) bare electrode, (c-d) ANT modified GR, and (e-f) after cycling at +0.3 V vs. Ag/AgCl in 0.1 M $H_2SO_4$.

For a practical purpose, we further modified discharged graphite rod (GR) isolated directly from low-cost and commercially available an Eveready dry cell (EV-GR, AA, 1V) via electrochemical grafting using anthracene diazonium salts (**Fig S32a**). Compared to unmodified EV-GR, ANT-modified EV-GR shows a Faradaic redox feature at ~ +0.36 V (vs. Ag/AgCl in 0.1 M $H_2SO_4$) and exhibits seven times increase in total areal capacitance. The large increase in capacitance observed with covalent ANT modifications calculated from GCD in − 0.1 V to + 0.7 V vs. Ag/AgCl at an applied current density of 30 µA/cm$^2$ for nine cycles from 615±7.5 µF cm$^{-2}$ to 9761±44 µF cm$^{-2}$ (**Fig S32b-d**). Further, the electrical conductivity of bare and modified GRs tested with two probes contact method and showed excellent conductivity of modified GR, which were further employed for lighting the LED using a power supply (**Fig S33-35**). All these experimental results strongly suggest the replacement of traditionally used graphitic rods with ANT-modified graphitic rods. Thus, our approach is highly encouraging to reuse of used carbon rods for energy applications; therefore, we developed a method of waste-to-energy conversion, a highly sustainable method. The anthracene-modified carbon electrodes can operate as excellent and high-performance electrochemical supercapacitors which originate from both contributions, non-Faradaic (electrochemical double layer) and Faradaic (bias-driven protonation/deprotonation) to various nitrogen fragments in molecular layers occurring in the positive potential ranges. The overall charging and discharging process are illustrated in **Fig. 6a-b**, which is mainly to the inclusion of proton ($H^+$) not only to pyridinic N, pyrolytic N, graphitic N, quaternary N, and N─N, but also to the porous surface of graphene rods. Though a similar protonation/deprotonation was described by Gan et al. and others but in the negative potential.[50–52] Anthracene having lower band gap (3.2 eV) over pyrene (3.4 eV) and naphthalene (4.3 eV, **Table S3**) is able to make films thicker than that of two, thus producing more nitrogen covering azo, which is the main sticking features of higher capacitance in anthrance-based molecular films. The authors have searched for positive materials, and our work can bridge the gap here.

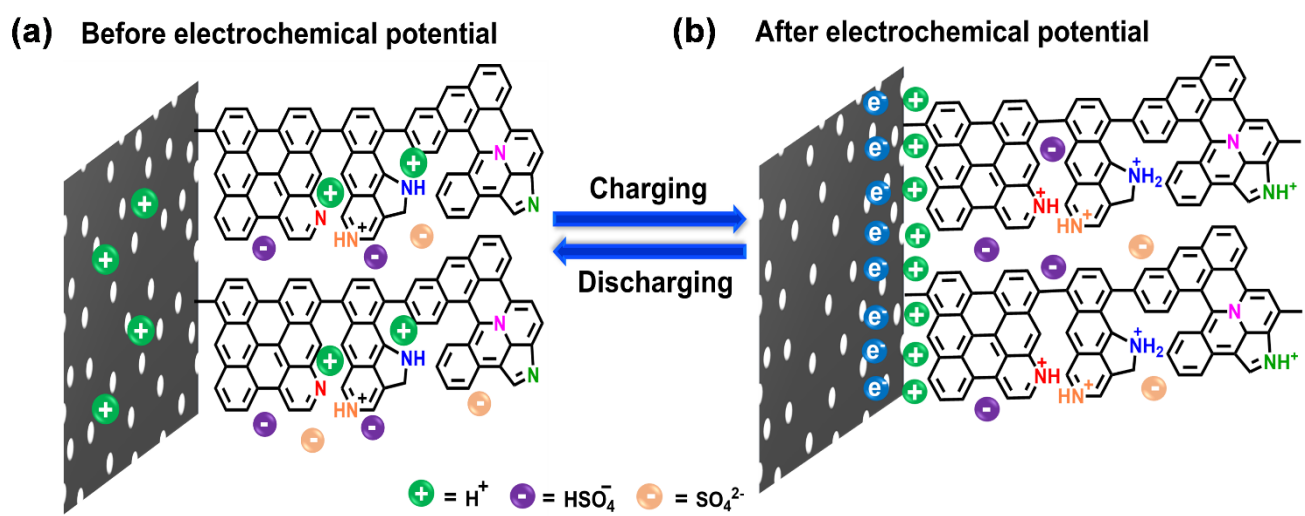

**Fig 6.** (a-b) Schematic illustration of charging and discharging with the bias-driven protonation/deprotonation to and from different nitrogen atoms present in the modified electrode.

## 3. Conclusion

As the world explores various novel possibilities for clean and safe energy sources, organic supercapacitors contribution is becoming more pronounced due to their environmentally friendlier, sustainable, and less dependence on inorganic metal oxide materials. Organic materials derived from naturally abundant materials (C, H, N, and O) are structurally diverse and can be functionally tailored by suitable synthetic ways to modify conducting carbon electrodes depending upon their applications, including in energy storage. Along with being based on naturally abundant elements, this chemistry makes it competitive with the inorganic electrodes when considering aspects such as environmental impact, sustainability, energy consumption, and cyclic stability. Also, the method demonstrated here has the bright potential to further extend the scope of functionalization on a variety of carbon electrodes, including flexible carbon materials, and can be employed for microfabricated flexible supercapacitor devices. To enhance our young family of organic supercapacitors, we demonstrate that low-cost graphite rods modification in nanoscale dimension via versatile electrochemical showing promising opportunity to reusages of waste carbon rods for large amount of energy storage applications. The present work opens new horizons in the field of "all-organic" nanoscale hybrid supercapacitors with baselines for future energy metrics to advance toward commercial implementations.

## 4. Experimental Section:

**Materials.** 1-aminoanthracene, 1-aminopyrene, 1-aminonaphthalene, sodium nitrite ($NaNO_2$), tertiary butyl nitrite, tetrafluoroboric acid ($HBF_4$, 48% in water), tetrabutylammonium perchlorate (TBAP) was purchased from Sigma-Aldrich (Bangalore, India), hydrochloric acid (Finar), acetonitrile (Avantore, HPLC-grade) was used as received. Cylindrical graphite rods (GR) used for electrochemical grafting were purchased from SPI supplies. Eveready dry cells were used to isolate graphite rods and cleaned with acetone for 1 hour and then with distilled water. Diazonium salts of all three amino precursors were synthesized by the reported methodology.

**E-chem grafting of different diazonium salts onto cylindrical graphite rods**

Prior to electrochemical modification, the graphite rod (GR) was pretreated with distilled water for 15 min, followed by drying with a nitrogen gun. The GRs were then dried in an oven at 100°C for 10 min to remove any traces of water present and then cleaned by a nitrogen gun to remove unwanted contaminants. E-chem grafting was performed employing a conventional three-electrode system using Metrohm Autolab Electrochemical workstation (Model: 204, software nova-2.14) and analyzed by using origin pro-9.1. The GR was used as a working electrode, $Ag/AgNO_3$ and Pt were used as reference and counter electrodes, respectively. All electrochemical measurements were recorded at room temperature. For electrochemical grafting, all diazonium solution concentrations were kept the same, that is 5 mM in HPLC-grade acetonitrile (ACN) with 0.1 M tetrabutylammonium perchlorate (TBAP) as a supporting electrolyte. Prior to E-Chem grafting, all diazonium solutions were deoxygenated by nitrogen purging for 15-20 min. After modification, the GR substrates were heavily rinsed with ACN to remove any physiosorbed materials.

**Characterization techniques**

All Synthesized diazonium salts were characterized by Bruker ALPHA II FTIR spectrometer and JASCO V-770 UV-Visible-NIR spectrophotometer (light source tungsten-deuterium lamp). Measurements were performed from 200 to 800 nm in a double beam mode and analyzed by origin pro-9.1. For thin film characterization, XRD was performed on bare GR and modified GR at room temperature in the 2θ range of 5° to 90° with PANanalytical Empyrean ACMS 101. Further, Raman spectroscopy was performed using SP2500i Acton instruments for monochromator and Spectrograph, laser source 532 nm from laser quantum. The surface morphology of the bare GR and electrochemically modified GR were analyzed by field emission scanning electron microscopy (FE-SEM) employing zeiss, model: supra40VP. In addition, X-ray photoelectron spectroscopy (XPS) was performed with Auger Electron Spectroscopy (AES) module PHI 5000 Versa Prob II, with a monochromator Al K-α source (1486.6 eV), hemispherical analyzer, and multichannel detector. Survey spectra of individual samples were obtained, followed by high-resolution spectra of C 1*s* and N 1s. $N_2$ adsorption-desorption measurement based on the Brunauer-Emmett-Teller (BET) principle was performed using Autosorb I; Quatachrome Corp. For thickness measurement, ANT-D was electrochemically grafted on an ITO substrate using similar grafting parameters. In tapping mode, surface morphology was analysed via atomic force microscopy (AFM). The AFM measurement was performed with Asylum Research (Oxford Instruments) with a silicon probe having a force constant of 0.5 N/m. The images were recorded at 0.60 Hz, set point in the range of 760 mV-730 mV with a drive frequency of 65.38 kHz. The thickness was evaluated by scratching the film with a sharp needle, and a line profile was drawn to get the thickness. All electrochemical Characterization was performed using a three-electrode system in 0.1 M $H_2SO_4$ that was purged with nitrogen for 15 min prior to all measurements employing above mentioned Autolab Electrochemical workstation. Ag/AgCl (saturated KCl) was used as a reference, and a platinum wire was used as a counter electrode. For constant current charge/discharge measurements, bare GR and modified GR substrates were charged from − 0.1 V vs. Ag/AgCl to + 0.7 V vs. Ag/AgCl and discharged from + 0.7 V vs. Ag/AgCl to − 0.1 V vs. Ag/AgCl to at different applied current densities. Solid-sate I-V measurements was performed using KEITHLEY 2604B source meter via two-probes contact.

**Conflicts of interest**

**Acknowledgments**

RG and AM thank IIT Kanpur for the senior research fellowship and post-doctoral fellowship, respectively. PCM acknowledges an initiation research grant from IIT Kanpur (IITK/CHM/2019044), and partial financial support from Science and Engineering Research Board (Grant No. CRG/2022/005325), and Council of Scientific & Industrial Research, project NO.:01(3049)/21/EMR-II, New Delhi, India.

## Table of content
**Nanoscale molecular electrochemical supercapacitors**

Ritu Gupta, Ankur Malik, Vincent Vivier* and Prakash Chandra Mondal*

The present work describes electrochemical modifications of graphite rods using aryl diazonium salts and provides a new perspective for the fabrication of environmentally friendly, sustainable, less-power consumption, and cyclic stable organic hybrid supercapacitors for future energy storage application.

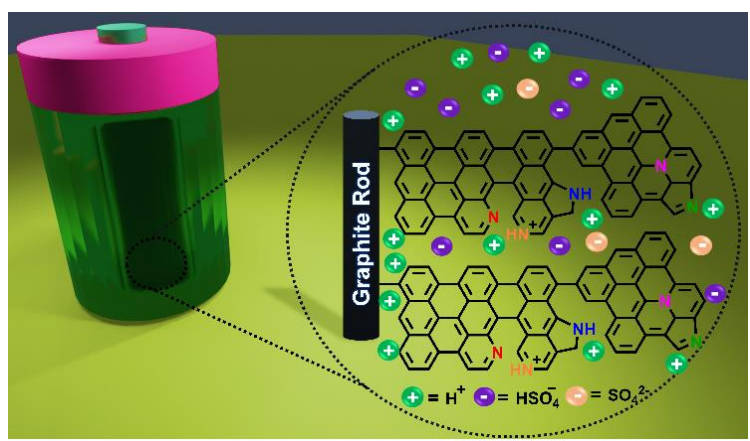